\documentstyle[aps,pre,epsfig]{revtex}

\begin{document}
\draft
\title{{\bf Correlation energy of an electron gas in strong magnetic fields at high densities }}
\author{M.Steinberg, J. Ortner}
\address{{\it Institut f\"ur Physik, Humboldt Universit\"{a}t zu Berlin, 
Invalidenstr. 110, D-10115 Berlin, Germany}}

\date{\today}
\maketitle
\begin{abstract}
The high-density electron gas in a strong magnetic field $B$ and at zero temperature is investigated. The quantum strong-field limit is considered in which only the lowest Landau level is occupied. It is shown that the perturbation series of the ground-state energy can be represented in analogy to the Gell-Mann Brueckner expression of the ground-state energy of the field-free electron gas. The role of the expansion parameter is taken by $r_B= (2/3 \pi^2)\,(B/m^2)\,(\hbar r_S /e)^3$ instead of the field-free Gell-Mann Brueckner parameter $r_s$. The perturbation series is given exactly up to $o(r_B)$ for the case of a small filling factor for the lowest Landau level.    
\end{abstract}


\section{Introduction}

The calculation of the correlation energy of a field-free and fully degenerate electron gas has a long history. Since the pioneering work of Gell-Mann and Brueckner \cite{Gell-Mann&Brueckner} and Wigner \cite{Wigner} who derived analytic results for the high-density and the low-density limit, respectively,  various interpolation formulas between these two limits have been established. For a free electron gas the ground-state energy is a function of the interelectron spacing $r_s$ only, which is related to the particle density $n$ by $n^{-1} =(4\pi r_s^3 a_B^3/3) $, where $a_B$ is the Bohr radius. Especially for a high-density electron gas the ground-state energy takes the form \cite{Gell-Mann&Brueckner}
\begin{equation}
\label{1} \epsilon_g = \frac{2.21}{r_s^2}-\frac{0.916}{r_s}+0.0622 \ln(r_s)-0.094 \, .  
\end{equation} 
The leading term is the kinetic energy. The next term is the first-order exchange energy, while the remaining terms in this series are called the correlation energy. \par
The purpose of our work is to find an analogous expansion of the ground-state energy for a system of a large number of electrons moving in a fixed uniform distribution of positive charge and in an external uniform magnetic field. In particular, we focus on the strong-field limit, where all the electrons are in the lowest Landau eigenstate and the spins are all aligned antiparallel to the magnetic field. The system is assumed to be at zero temperature. In our analysis we essentially follow the calculation originally developed by Gell-Mann and Brueckner. A homogeneous magnetic field $B$ modifies the energy spectrum of a charged particle and the Fermi energy (i.e., the chemical potential of the noninteracting magnetized electron gas) is no longer a function of $r_s$ alone, but also of the magnetic field. Therefore, it is convenient to introduce the new parameters 
\begin{equation}
\label{0.20} r_B = \frac{1}{\pi a_B k_F} =\frac{2}{3\pi^2} \alpha^2 r_s^3 \, \, , \, \hspace{0.5cm}\, \hspace{0.5cm} t=\frac{\epsilon_F}{\hbar \omega_c}=\frac{9\pi^2}{8} \frac{1}{\alpha^6 r_s^6} \, ,
\end{equation}
where $\omega_c=eB/m$ is the cyclotron frequency, $\alpha=a_B/l_B$ is the ratio of the Bohr radius, and the magnetic length $l_B=\sqrt{\hbar/(eB)}$. At strong magnetic fields $r_B$ takes the role of the expansion parameter. The second parameter $t$ may be regarded as a filling parameter. It must satisfy the condition $t\leq 1$ in order to describe a quantum system in the strong-field limit. In both equations we have made use of the the relation between the Fermi wave vector $k_F$ and the particle density in the strong-field limit, where the system behaves essentially as a one-dimensional electron gas  
\begin{equation}
\label{0.0} n=\frac{1}{(2\pi l_B)^2} \int_{-k_F}^{k_F} dk_z \, , \hspace*{1cm} k_F=2\pi^2 l_B^2 n \, .  
\end{equation}
Now the kinetic energy of the noninteracting degenerate electron gas per particle in rydbergs becomes
\begin{equation}
\label{0.1} \epsilon_{kin}= \frac{1}{(2\pi l_B)^2} \left(\int_{-k_F}^{k_F} \frac{\hbar^2 k_z^2}{2m} \, dk_z\right) \frac{1}{n \, Ryd} =\frac{1}{3\pi^2} \frac{1}{r_B^2} \approx \frac{0.0337}{r_B^2} .
\end{equation}
Thus the kinetic energy is a function of the dimensionless parameter $r_B$ only. In the limit $r_B\ll 1$, the kinetic energy will give the dominant contribution to the ground-state energy, while the Coloumb interaction acts as a small perturbation to the motion of the electrons. We investigate here this physically interesting high-density regime. We show that in this case the ground-state energy takes the form  

\begin{equation} 
\label{0.3} \epsilon_{g} = \frac{1}{3\pi^2} \frac{1}{r_B^2}+\frac{A(t)}{r_B}+B(t) \ln(r_B) + C(t)+ {\rm terms \, \, that \, \, vanish \, \,  as \, \, r_B \rightarrow 0} \, .
\end{equation}
The second term in this series comes from the first-order exchange energy. Its calculation was given by Danz and Glasser and will be reviewed in Sec. II. The terms in the energy beyond the Hartree-Fock approximation can be obtained by a formal summation of all ring diagrams. In addition to that, we must include the second-order exchange diagram that contributes to $C(t)$. Depending on the values of the magnetic field and the density, the filling parameter $t$ may vary within the range $0 < t \leq 1$. An analytical calculation for the coefficients $A(t)$, $B(t)$ and $C(t)$ is too difficult as it involves a troublesome summation over Landau indices. Therefore, we give explicit expressions for the contributions to the constants $A(t)$, $B(t)$, and $C(t)$ and find analytic results for their asymptotic behavior as $t$ goes to zero. The validity domain of this expansion in the $r_s$-$t$ plane is schematically shown in Fig. 1, in which we have made use of the relationship $r_B \sim r_s/t^{\frac{1}{3}}$ to indicate the high-density region ($r_B\ll1$). The regime of a coupled electron gas ($r_B>1$) was investigated by Fushiki et al \cite{Fushiki92} using the Thomas-Fermi type statistical model. At sufficiently low densities the electron gas should form a Wigner crystal. Following Kleppmann and Elliott \cite{Kleppmann&Elliot} (see also Ref. \cite{MacDonald}), a strong magnetic field would increase the density at which crystallization occurs. The results obtained in this work may be used to establish interpolation formulas between the various limiting results. 
\begin{figure}[h]
\centerline{\epsfig {figure=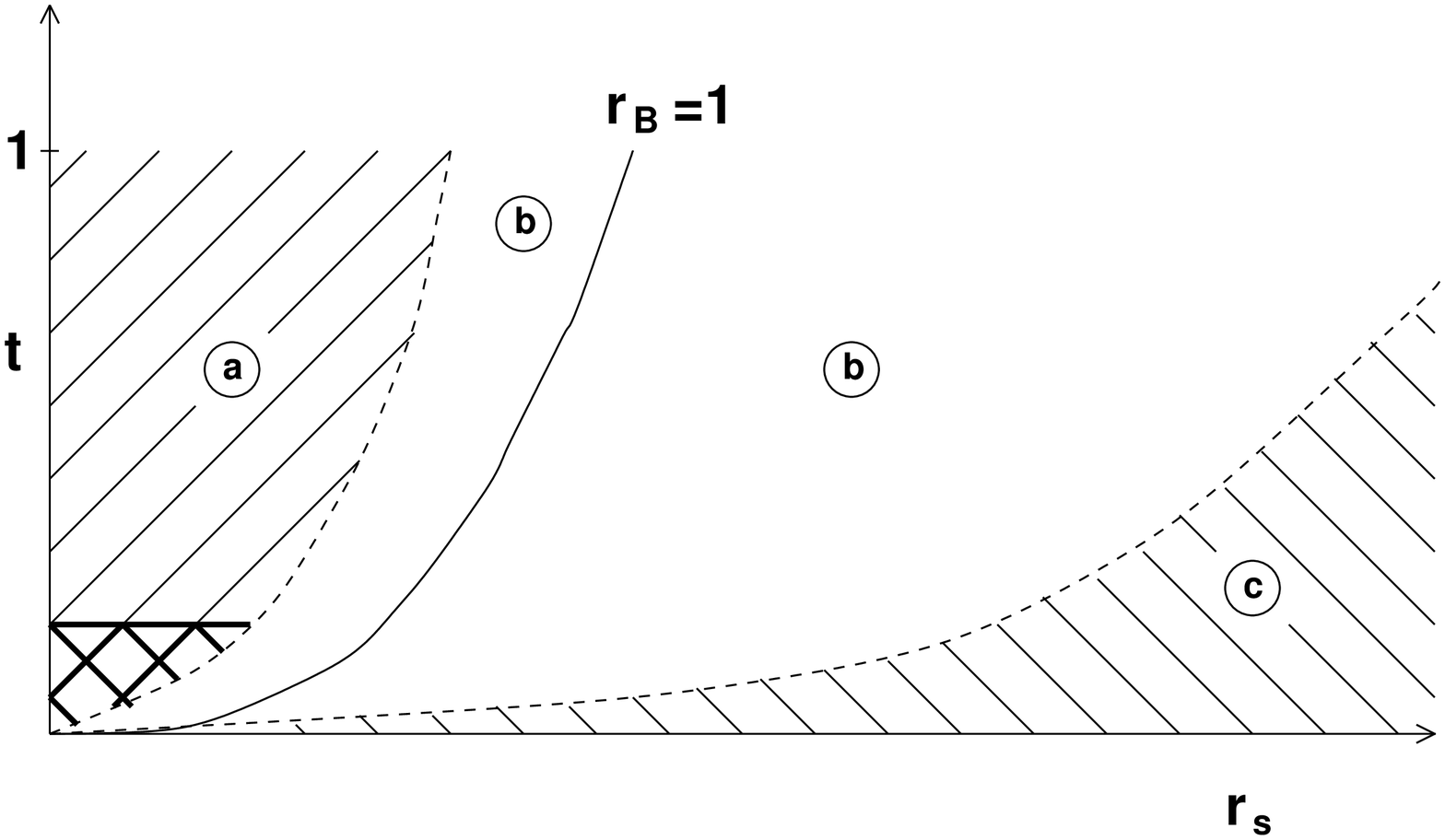,width=9.4cm,angle=0}}
\caption{\sl Scetch of the validity domain of the various expansions in the $r_s-t$ plane: (a) high-density region [the cross-hatched area at the bottom shows the validity domain of the analytic results for the asymptotic behavior $t \rightarrow 0$ of the coefficients  $A(t)$, $B(t)$ and $C(t)$.], (b) intermediate-density region, and (c) low-density region. }
\end{figure}

The strong field limit has also been investigated by Horing et al. \cite{Horing&Danz&Glasser} and Isihara and Tsai \cite{Isihara}. We have partially adopted the notation of Horing et al. and additionally we have introduced the parameter $t$ to clarify the structure of the perturbation expansion of the ground-state energy Eq.(\ref{0.3}). This was not explicitly given in Ref. \cite{Horing&Danz&Glasser}. Horing et al. \cite{Horing&Danz&Glasser} carefully analyzed the spectrum of the plasma oscillations  within the random-phase approximation to calculate the correlation energy. It is found to be proportional to $\ln(r_B)$. We show that the same result can be obtained by expanding the polarization function in powers of the momentum transfer, which gives a better understanding of the structure of the ground-state energy expansion and simplifies the calculation. Furthermore, this enables us to find the constant $C(t)$ of this series. This will be discussed in Sec. III.    

The calculations of Isihara and Tsai \cite{Isihara} are performed in the grand canonical ensemble. Within our notations their ground-state energy agrees with that of Horing et al. \cite{Horing&Danz&Glasser} up to the order $A(t)/r_B$, but differs in the constant $B(t)$ by a factor 2. We confirm in our calculation the result of Horing et al. \cite{Horing&Danz&Glasser}. Isihara and Tsai \cite{Isihara} also give contributions to the term $C(t)$; however, within this order their calculation is not complete.  

Our work is motivated by the observation that all of these conditions [degeneracy, weak coupling ($r_B\ll 1$) and small filling factor ($t \ll 1$)] may be realized on the surface of some strongly magnetized neutron stars \cite{Lai&Salpeter3}. The outermost layers of these objects are likely to form a degenerate hydrogen plasma at magnetic field strengths of $10^9 T$, at temperatures of $10^5 K$ and at densities of $3 \times 10^{34} m^{-3}$. Then the expansion parameter and the filling factor are found to be $r_B \approx 0.01$ and $t \approx 0.05$. On the other hand, at higher temperatures ($T>10^5 K$) and lower densities ($n<10^{30} m^{-3}$) the magnetized plasma ($B=10^9 T$) is nondegenerate and weakly coupled. This situation has been recently considered in Refs. \cite{Steinberg&Ortner&Ebeling} and \cite{Potekhin}. 

The high-density strong-field limit may also be achieved in laboratory plasmas, such as semiconductors or laser-induced plasmas. In particular, one may consider semiconductors with small effective mass $m^\star=0.01 m_e$ and carrier densities of $5 \times 10^{23} m^{-3}$ in strong magnetic fields of $40 T$, such as indium antimonide.  At these values of $B$ and $n$ the parameters $r_B \approx 0.02$ and $t\approx 0.04$ lie within our approximation scheme. Throughout this paper the energies are given per particle and in rydbergs. 

\section{Exchange energy}

In this section we shall briefly review the calculation of the first-order exchange energy, which is diagrammatically represented in Fig. 2.

\begin{figure}[h]
\centerline{\epsfig {figure=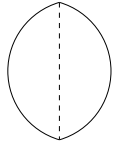,width=1.4cm,angle=0}}
\caption{\sl First order exchange diagram.}
\end{figure}

This contribution has been examined by Danz and Glasser \cite{Danz&Glasser}. Following the rules of calculating Feynman diagrams we have 

\begin{equation}
\label{0.16} \epsilon_{HF} = \frac{1}{2n \, Ryd} {\bf Tr}_{\sigma} \int_0^1 \frac{d\lambda}{\lambda}   \int d1 d2 \, V^\lambda(12) \, G^<_\sigma(12) \, G^<_\sigma(21) \, .
\end{equation}
The trace is over the spin variable $\sigma_z = (-1,+1) $. Throughout this work we shall use the closed form of the Green's function $G^<_\sigma(12)$ in space-time representation $ 1=({\bf r}_1,t_1)$ as given by Horing \cite{Horing}, which has also been exploited in Refs.\cite{Steinberg&Ortner&Ebeling} and \cite{Danz&Glasser}  

\begin{eqnarray}
\label{0.17} G^{\{ {> \atop <} \}}_\sigma(12) & = & C({\bf r,r^\prime}) \int \frac{d\omega}{2\pi} \,   {\Bigg\{} {-i[1-f_0(\omega)]   \atop if_0(\omega)} {\Bigg\}} \, e^{-i\omega T}  \int_{-\infty}^\infty dT^\prime \,  \int \frac{d{\bf p}}{(2\pi)^3} \, e^{i{\bf p} ({\bf r}_1-{\bf r}_2)} \, \exp\left[ -i\left(\mu_B B\sigma_z+\frac{p_z^2}{2m}-\omega\right) T^\prime \right]   \nonumber\\
\label{0.18} & & \times \frac{1}{\cos\left(\frac{\omega_c}{2} T^\prime\right)} \, \exp\left[ -i \frac{p_x^2+p_y^2}{m\omega_c} \tan\left(\frac{\omega_c}{2}T^\prime\right) \right] \, . 
\end{eqnarray}
Here we have introduced the Fermi-Dirac distribution function $f_0(\omega)$, the Bohr magneton $\mu_B=e\hbar/(2m)$ and a unitary phase factor $C({\bf r,r^\prime})$. After inserting expression (\ref{0.18}) into Eq.(\ref{0.16}) and considering the quantum strong-field limit only, i.e., $ \hbar \omega_c > \epsilon_F$, one can perform all elementary integrals to obtain the Hartree-Fock exchange energy  
\begin{equation}
\label{0.2} \epsilon_{HF}  =  - \frac{1}{r_B} \frac{4}{\pi^2} \int_0^\infty dp \left[\arctan \left(\frac{1}{p}\right) -\frac{p}{2} \ln{\left(1+\frac{1}{p^2}\right)} \right] \exp(-4tp^2) \, . 
\end{equation}
In the limit $t \rightarrow 0$ the exchange energy takes the simple form 
\begin{equation}
\label{0.32} \epsilon_{HF} (t \rightarrow 0) =  - \frac{1}{r_B} \, \frac{1}{\pi^2} \left(3-{\bf C} - 2\ln(2)-\ln(t) \right) \, , 
\end{equation}
where ${\bf C}$ is Euler's constant ${\bf C} \approx 0.5772$. This result was already obtained by Danz and Glasser \cite{Danz&Glasser}. One can easily identify the coefficient $A(t)$ from Eq.(\ref{0.2}) and its asymptotic limit $t\rightarrow 0$ from Eq.(\ref{0.32}). 

\section{Calculation of the correlation energy}

In second-order perturbation theory, one expects a higher-order contribution in the interaction parameter $e^2$ or $r_B$, that is, a constant independent of $r_B$. Like in the zero-magnetic-field counterpart, this contribution diverges logarithmically at small momentum transfers. As was shown by Gell-Mann and Brueckner, this difficulty can be overcome by summing ring diagrams up to infinite order (Fig. 3), which is known as the random-phase approximation.
\begin{figure}[h]
\begin{equation}
\hspace{2.2cm}  \epsilon_r = \hspace{0.6cm}    \begin{minipage} [htbp]{10cm} \vspace*{0.65cm} \psfig{figure=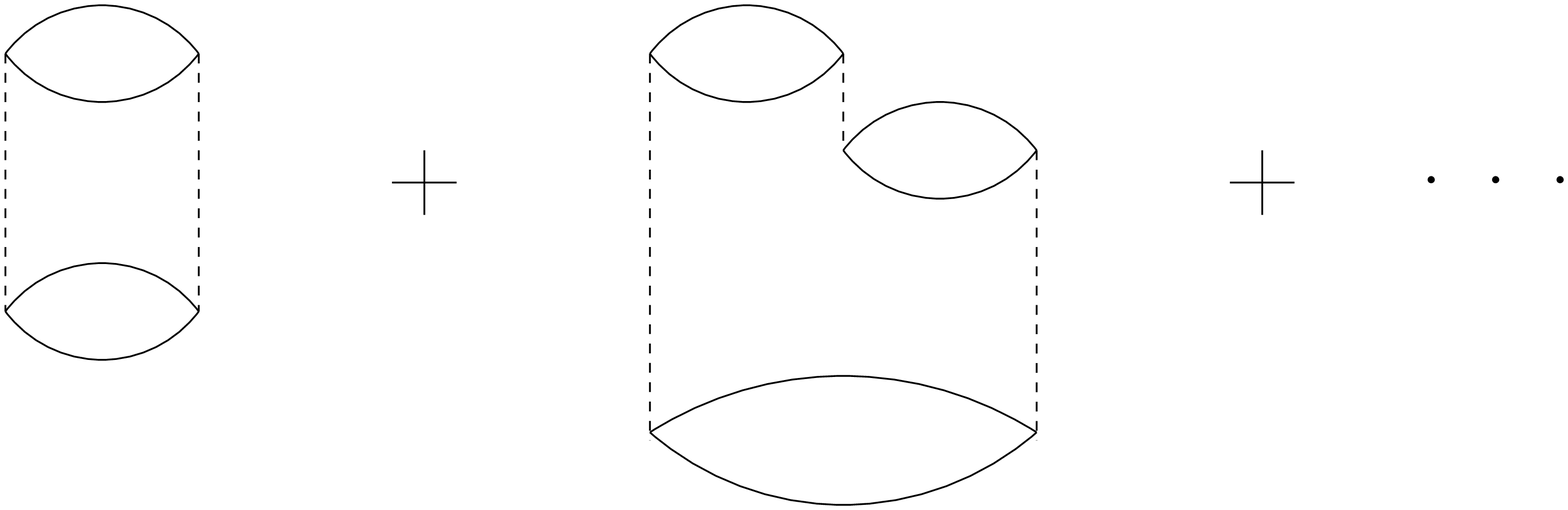,width=8.4cm,angle=0} \end{minipage} 
\end{equation}
\caption{\sl Ring approximation for the correlation energy.}
\end{figure}
Given the polarization function $\Pi({\bf  p},\omega)$ within the random-phase approximation, the correlation energy may be written as
\begin{equation}
\label{0.4} \epsilon_{r} =  \frac{\hbar k_F^3}{2 n \, Ryd} \int \frac{d{\bf p}}{(2\pi)^3} \int \frac{d\omega}{(2\pi)} \left[ \ln\left(1+\frac{e^2}{{\bf p}^2} \Pi({\bf  p},\omega)\right) - \frac{e^2}{{\bf p}^2} \Pi({\bf  p},\omega)\right] \, .
\end{equation}
Here and in the remainder of this paper we have expressed all momenta in terms of the Fermi wave vector $k_F$. For a further detailed analysis, the polarisation function $\Pi({\bf  p},\omega)$ must be evaluated. This quantity is determined solely by a collisionless, i.e., noninteracting, Fermi system in a magnetic field and may be calculated in a space-time representation by
\begin{equation}
\label{0.19} \Pi(12) = i \hbar^{-1}  {\bf Tr}_\sigma \, \left(G_\sigma(12) G_\sigma(21) \right) \, .
\end{equation}
By using Eq.(\ref{0.18}), the polarization function was studied in detail by Horing \cite{Horing}. He obtained useful analytical results for the nondegenerate as well as for the degenerate region. In particular he derived an exact expression for the polarization function in the quantum strong-field limit [Ref. \cite{Horing}, p.61, Eq.(A.III.6)), which is given by  
   
\begin{equation}
\label{0.5} \Pi({\bf  p},\omega) = - \frac{r_B}{2e^2 t}   \frac{1}{{|p_z|} } e^{(-p_\rho^2 t)} \sum_{n=0}^\infty \frac{1}{n!} \left(p_\rho^2 t\right)^n \ln\left(\frac{\left(\frac{m\omega}{\hbar k_F^2}\right)^2+\left[\frac{n}{2t}+\frac{p_z^2}{2}-| p_z | \right]^2}{\left(\frac{m\omega}{\hbar k_F^2}\right)^2+\left[\frac{n}{2t}+\frac{p_z^2}{2}+| p_z | \right]^2} \right).
\end{equation}
Here we have introduced our notation [Eq.(\ref{0.20})]. We have also made use of Eq.(\ref{0.0}), which relates the Fermi wave vector $k_F$ to the expansion parameter $r_B$. Equations (\ref{0.4}) and (\ref{0.5}) represent the basis of our further calculations. From the ring approximation $\epsilon_r$ we may find the contribution to $B(t)$ and $C(t)$ by expanding the polarization function in powers of the momentum transfer ${\bf p}$. Only the lowest-order term independent of ${\bf p}$ must be considered, while higher-order terms proportional to ${\bf p}^2$ will give higher powers of $r_B$ in the final expression of the ground-state energy. An additional contribution to $C(t)$ arises in the exchange interaction of the order $e^4$. This energy may be represented diagrammatically as in Fig. 4. 
\begin{figure}[h]
  \centerline{ \epsfig{figure=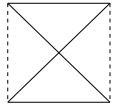,width=1.4cm,angle=0}}  
\caption{\sl Second-order exchange diagram.}
\end{figure}
Higher-order exchange terms lead to contributions in higher powers of $r_B$ and need not be considered in our calculation. 
\subsection{ring approximation}
We may start our calculation with the investigation of the ring approximation. Following the original work of Gell-Mann and Brueckner, we first establish an expansion of the polarization function in powers of the transfer momentum. For this we introduce the dimensionless frequency $ u \rightarrow (m\omega)/(\hbar k_F^2 |p_z|) $. Now we fix this quantity and let the momentum transfer approach zero to obtain the limiting form 
\begin{equation} 
\label{0.8} \frac{2e^2t}{r_B} \times  \Pi({\bf  p},u) \approx R_1(u) +  p^2 R_2(u, \cos(\theta)) + ...
\end{equation}
In general the functions $R_1$, $R_2$, etc., will depend on the direction of the transferred momentum with respect to the magnetic field that is taken into account by the additional argument $\cos(\theta)$ in $R_2$. However, as may be easily shown, $R_1$ is independent of this angle and takes the simple form     
\begin{equation} 
\label{0.9} R_1(u) =\frac{2}{1+u^2} \, ,
\end{equation}
Now we perform the integral over $p$ in Eq. (\ref{0.4}) from $0$ to 1 and correct to order $r_B$ exactly. This may be achieved by introducing $\delta(t)$, which restores the second-order pertubation term to its exact value. Thus we have   
\begin{equation} 
\label{0.10} \epsilon_{r} \approx \frac{1}{2 \pi^3} \frac{t}{r_B^2} \int_0^1 dp \, p^3 \int_{-\infty}^\infty  du \left[ \ln{\left(1+\frac{r_B}{2t} \frac{1}{p^2} R_1(u) \right)}  - \frac{r_B}{2t} \frac{1}{p^2} R_1(u)   \right] + \delta(t) \, ,
\end{equation}
where $\delta$ is defined by
\begin{equation}
\label{0.11} \delta(t) = \epsilon_c^{(2)} +\frac{1}{8\pi^3 t}  \int_0^1 dp \int_{-\infty}^\infty du \frac{R_1^2}{2p} \, .
\end{equation}
This is a finite expression since the logarithmic divergence cancels. Substituting Eq. (\ref{0.5}) into this equation, one obtains the general form of this $\delta(t)$ as given in Appendix A [Eq.(\ref{app1.0.1})]. There we have derived an asymptotic expansion that is found to be 
\begin{equation}
\label{0.34} \delta (t\rightarrow 0) = -\frac{1}{16 \pi^2 t}-\frac{\ln(t)}{6}-\frac{\left( {\bf C} +3 \ln(2)\right)}{6} +\frac{1}{3}-\frac{4}{\pi^2} \zeta(3) \, ,
\end{equation}
where we have introduced the Rieman zeta function $\zeta$ with $\zeta (3) \approx 1.202 057 $. The integration over the momentum in Eq. (\ref{0.10}) is readily carried out with the result 
\begin{equation}
\label{0.12} \epsilon_{r} \approx \frac{1}{32\pi^3t} \int_{-\infty}^\infty du (R_1(u))^2 \left[\ln{\left( \frac{r_B}{2t} R_1(u)\right)} - \frac{1}{2}\right] \, ,
\end{equation}
where we have dropped all terms that vanish as $r_B \rightarrow 0$. The remaining integral may be performed exactly and we find for the logarithmic contribution, i.e., for $B(t)$, the result
\begin{equation} 
\label{0.13} B(t)=\frac{1}{16\pi^2t} \, .
\end{equation}
Note that $B(t)$ is exact for arbitrary filling factor. This result is in agreement with that of Horing et al \cite{Horing&Danz&Glasser}.  Furthermore, we are able to determine the next term in the perturbation series. Again, we can perform the integral (\ref{0.12}) and take into account the contributions coming from $\delta(t)$ and $\epsilon_{ex}(t)$ to obtain 
\begin{equation} 
\label{0.14} C(t)=\frac{1}{16\pi^2t} \left(\frac{1}{2}- 2\ln(2)-\ln(t) \right) + \delta (t) +\epsilon_{ex}(t) \, .
\end{equation}
We may proceed with the calculation of the second-order exchange term in the next subsection.
\subsection{second-order exchange term} 
This contribution may be written in space-time representation as
\begin{equation}
\label{0.6} \epsilon_{ex} (t) =  \frac{1}{2n \, Ryd} {\bf Tr}_{\sigma} \int_0^1 \frac{d\lambda}{\lambda} \int d1 \int d2 \int d3 \int d4 V^\lambda(14) V^\lambda(23) G_{\sigma}(12) G_{\sigma}(23) G_{\sigma}(34) G_{\sigma}(41) \, .
\end{equation}
Now we may use Eq.(\ref{0.17}) and employ the expansion of the Green's function in terms of the Laguerre polynomials $L_s(x)$ [Ref. \cite{Danz&Glasser}, p.96, Eq.(2.3)]. Then the $T^\prime$ and $\omega$ integrations are trivial and in the limit $\hbar \omega_c > \epsilon_F$ this contribution can be written as 
\begin{eqnarray}
\epsilon_{ex} (t) & = & \frac{4 n^2 \l_B^6}{\pi} \sum_{s=0}^\infty \sum_{s^\prime=0}^\infty \int_{\left((p_z-q_z)^2+\frac{s}{t} \right)>1 \atop{\left((q_z+k_z)^2+\frac{s^\prime}{t} \right)>1} } d{\bf q} \int_{|p_z|<1} d{\bf p} \int_{|k_z|<1} d{\bf k} \frac{1}{{\bf q}^2} \frac{1}{|{\bf q}+{\bf k}-{\bf p}|^2} \frac{1}{q_z (q_z+k_z-p_z)+\frac{(s+s^\prime)}{2t}} \nonumber\\
&&  \times \exp\left[-2t\left( {\bf p}_\rho^2 + {\bf k}_\rho^2+({\bf p}_\rho-{\bf q}_\rho)^2+({\bf q}_\rho+{\bf k}_\rho)^2\right) \right] (-1)^{s+s^\prime} {\bigg(}L_s \left[4t ({\bf p}_\rho-{\bf q}_\rho)^2\right] L_{s^\prime}  \left[4t ({\bf q}_\rho+{\bf k}_\rho)^2\right] \nonumber\\
\label{0.21} && + \Theta(s-1) \Theta(s^\prime-1) L_{s-1} \left[4t ({\bf p}_\rho-{\bf q}_\rho)^2\right]  L_{s^\prime-1}  \left[4t ({\bf q}_\rho+{\bf k}_\rho)^2\right] {\bigg)} \, .
\end{eqnarray}
Here ${\bf p_\rho}$ denotes the two-dimensional vector ${\bf p_\rho}=(p_x,p_y)$. This expression is the basis for the caculation of $C(t)$ at arbitrary filling factor $t<1$. For small values of $t$ we may seek an asymptotic exact result for $\epsilon_{ex} (t)$. We find that in this limit only the state $s=s^\prime=0$ contributes to the exchange energy, while the other terms ($s\geq1$ or $s^\prime \geq 1$) vanish as $t$ goes to zero. Thus we have, for $t\ll 1$,

\begin{eqnarray}
\label{0.7} \epsilon_{ex} (t) & = & \frac{2t}{\pi^5} \int_{|p_z-q_z|>1 \atop{|q_z+k_z|>1} } d{\bf q} \int_{|p_z|<1} d{\bf p} \int_{|k_z|<1} d{\bf k} \frac{1}{{\bf q}^2} \frac{1}{|{\bf q}+{\bf k}-{\bf p}|^2} \frac{1}{q_z (q_z+k_z-p_z)} \nonumber\\
&&  \times \exp\left[-2t\left( {\bf p}_\rho^2 + {\bf k}_\rho^2+({\bf p}_\rho-{\bf q}_\rho)^2+({\bf q}_\rho+{\bf k}_\rho)^2\right) \right] \, .
\end{eqnarray}
The detailed evaluation of this integral in the asymptotic limit is given in Appendix B. It shows an interesting logarithmic dependence on the filling factor, which is given by  
\begin{eqnarray} 
 \epsilon_{ex} (t\rightarrow 0) & = & \frac{1}{6} \ln^2(t)+\left(-\frac{2}{3}+\frac{{\bf C}}{3}+\ln(2) +\frac{8 \zeta(3)}{\pi^2}\right) \ln(t) +\frac{4}{3}-\frac{2{\bf C}}{3}+\frac{{\bf C}^2}{6} +\frac{13\pi^2}{90} -2 \ln(2)+\frac{3 \ln^2(2)}{2}\nonumber\\
\label{0.22} & & + {\bf C} \ln(2) +\frac{\zeta(3)}{\pi^2} \left(24 \ln(2)+8 {\bf C} -16 \right)\, .
\end{eqnarray}

\section{conclusions} 
In this paper the ground-state energy of a degenerate electron gas in a strong magnetic field has been investigated. The electron gas in the strong-field limit is characterized by two parameters: the dimensionless inverse Fermi wave vector $r_B$ and the filling factor of the lowest Landau level $t$ [see Eq.(\ref{0.20})]. We have shown that in the case of a dense electron gas $r_B \ll 1$ the internal energy per particle (measured in rydbergs) may be expressed in the form of Eq. (\ref{0.3}). The constant $A(t)$ is fully determined by the Hartree-Fock exchange energy and may be identified by using Eq.(\ref{0.2}). The next term in this expansion comes from the random-phase approximation correlation energy. The constant $B(t)$ of the $\ln(r_B)$ contribution is given by Eq.(\ref{0.13}). For the calculation of the constant $C(t)$ one needs to include the second-order exchange energy. Given $\delta(t)$ and $\epsilon_{ex}(t)$ by Eq.(\ref{app1.0.1}) and Eq.(\ref{0.21}), respectively, one may calculate the constant $C(t)$ using Eq.(\ref{0.14}) at any $t$. 

For the case of very strong magnetic fields with a small filling factor t (but sufficiently small $r_B$) we have given analytic expressions for the constants $A(t)$ by Eq.(\ref{0.32}) and $B(t)$ by Eq.(\ref{0.13}). Our results for $A(t \rightarrow 0)$ and $B(t \rightarrow 0)$ coincide with earlier calculations of Danz and Glasser \cite{Danz&Glasser} and of Horing et al. \cite{Horing&Danz&Glasser}. The calculation of the constant $C(t \rightarrow 0)$ is, to our knowledge, given for the first time. We can use the asymptotic result for $\delta(t) $ [Eq.\ref{0.34}] and for $\epsilon_{ex} (t) $ [Eq.\ref{0.22}] to get an expression for small values of $t$:
\begin{eqnarray} 
 C(t\rightarrow 0) & = & -\frac{\ln(t)}{16\pi^2t} - \frac{2\ln(2)+\frac{1}{2}}{16\pi^2 t}+\frac{\ln^2(t)}{6}+\left(-\frac{5}{6}+\frac{{\bf C}}{3}+\ln(2)+\frac{8\zeta(3)}{\pi^2}\right) \ln(t)  \nonumber\\
\label{0.15}&& + \frac{5}{3}-\frac{5{\bf C}}{6}+\frac{{\bf C}^2}{6}+\frac{13 \pi^2}{90}+{\bf C} \ln(2) -\frac{5\ln(2)}{2}+\frac{3\ln^2(2)}{2}+\frac{\zeta(3)}{\pi^2} \left(24 \ln(2)+8 {\bf C}-20\right)\, . 
\end{eqnarray}
One should emphasize that Eq.(\ref{0.3}) with the asymptotic results for the constants [Eqs.(\ref{0.32}),(\ref{0.13}), and (\ref{0.15})] is an asymptotic expansion of the ground-state energy of an electron gas in the limit $r_B \rightarrow 0$ and then $t \rightarrow 0$ with a possibly poor convergence. Including the results of our earlier work \cite{Steinberg&Ortner&Ebeling}, where the thermodynamic perturbation expansion in the Boltzmann limit has been performed, we have now found exact results of the thermodynamic functions in the limiting cases of a degenerate and nondegenerate weakly coupled electron gas. A third known limiting case is the region of Wigner cristallization (see Refs. \cite{Kleppmann&Elliot} and \cite{MacDonald} and references therein). One may now try to connect the known limiting results by a Pad\'{e} approximation in order to describe the intermediate region as indicated in Fig. 1. A survey of several Pad\'{e} approximations for the case of the zero-field electron gas was recently given by Stolzmann and Bl\"ocker \cite{Stolzmann&Bloecker}. A Pad\'{e} approximation for the thermodynamic functions of the electron gas in the quantum strong-field limit will be given elsewhere.  

{\section[Acknowledgments]{Acknowledgments}
Valuable discussions with Werner Ebeling are gratefully acknowledged. This work was supported by the Deutsche Forschungsgemeinschaft under grant No. Eb 126/5-1.

\begin{appendix}
\section{Calculation of $\delta$}  
This contribution may be written in terms of the polarizationfunction (\ref{0.5}) as
\begin{eqnarray}
\delta (t) & = & - \frac{1}{32\pi^4t} \int d{\bf p} \int_{-\infty}^\infty d\omega \frac{1}{{\bf p}^4} \frac{e^{-2p_\rho^2t}}{|p_z|^2} \sum_{n,m} \frac{1}{n!m!} \left( p_\rho^2t\right)^{n+m} \ln\left(\frac{\omega^2+\left[\frac{n}{2t}+\frac{p_z^2}{2}-| p_z | \right]^2}{\omega^2+\left[\frac{n}{2t}+\frac{p_z^2}{2}+| p_z | \right]^2} \right)  \nonumber\\
\label{app1.0.1} & \times & \ln\left(\frac{\omega^2+\left[\frac{m}{2t}+\frac{p_z^2}{2}-| p_z | \right]^2}{\omega^2+\left[\frac{m}{2t}+\frac{p_z^2}{2}+| p_z | \right]^2} \right) + \frac{1}{8\pi^3 t}  \int_0^1 dp \int_{-\infty}^\infty d\omega \frac{R_1^2}{2p} \, .
\end{eqnarray}
We are interested in small values of $t$, therefore we can restrict the calculation to $n=m=0$ as $n$ or $m\geq 1$ give rise to higher-order terms in $t$. Note that the terms  $n$ or $m\geq 1$ are convergent expressions even without subtracting the logarithmic divergent part, while the lowest-order term $n=m=0$ would show singular behavior. The $\omega$ integration in Eq. (\ref{app1.0.1}) may be performed exactly. The result may be split into small- and large-wave-number contributions $\delta (t) = \delta_1 (t) + \delta_2 (t) $ with 
\begin{equation}
\label{app1.0.2} \delta_1 (t) = - \frac{1}{8\pi^2 t} \int_0^1 dp \int_0^1 dz \frac{e^{-8tp^2 \left(\frac{1}{z}-1\right) }}{p^4}  \left[(p-p^2)\ln(p-p^2)+(p+p^2)\ln(p+p^2) -2p\ln(p) \right] + \frac{1}{8\pi^2 t} \int_0^1 dp \frac{1}{p}
\end{equation}
and
\begin{equation}
\label{app1.0.3} \delta_2 (t) = - \frac{1}{8\pi^2 t} \int_1^\infty dp \int_0^1 dz \frac{e^{-8tp^2 \left(\frac{1}{z}-1\right) }}{p^4}  \left[(p^2-p)\ln(p^2-p)+(p^2+p)\ln(p^2+p) - 2p^2\ln(p^2) \right] \, .
\end{equation}
First we focus on the calculation of $\delta_1(t)$. The z integration may be expressed in terms of the exponential integral Ei(-x) 
\begin{equation}
\label{app1.0.4} \delta_1 (t) = - \frac{1}{8\pi^2 t} \int_0^1 dp \frac{1}{p^3} \left(1+8tp^2 e^{8tp^2} Ei(-8tp^2)\right)  \left((1-p)\ln(1-p)+(1+p)\ln(1+p)\right) +\frac{1}{8\pi^2 t} \int_0^1 dp \frac{1}{p} \, .
\end{equation}
Now we can expand the logarithm and subtract the divergent part from the lowest-order contributions. Thus we have
\begin{equation}
\label{app1.0.5} \delta_1 (t) = - \frac{1}{8\pi^2 t} \int_0^1 dp \left\{ \sum_{k=2}^\infty \frac{2}{2k (2k-1)} p^{2k-3} + 8t \sum_{k=1}^\infty \frac{2}{2k (2k-1)} p^{2k-1} e^{8tp^2} Ei(-8tp^2) \right\} \, .
\end{equation}
Again, we can expand the exponential and the exponential integral and retain only contributions that do not vanish as $t\rightarrow 0$, 
\begin{equation}
\label{app1.0.6} \delta_1 (t\rightarrow 0) = - \frac{1}{8\pi^2 t} \sum_{k=2}^\infty \frac{2}{2k (2k-1) (2k-2)} - \frac{1}{\pi^2}  \sum_{k=1}^\infty \left\{  \frac{2 ({\bf C}+\ln(8t))}{(2k)^2 (2k-1)} - \frac{4}{(2k)^3 (2k-1)} \right\} \, .
\end{equation}
This may be simplified by expressing the sums in terms of the $\psi$ function and observing that they give elementary constants for integer arguments. Therefore, the final result may be written as
\begin{equation}
\label{app1.0.12} \delta_1 (t\rightarrow 0) = \frac{4\ln(2)-3}{16\pi^2 t } + \frac{2\left(\frac{\pi^2}{24}-\ln(2)\right)}{\pi^2} \ln(t) +\frac{2\left(\frac{\pi^2}{24}-\ln(2)\right) ({\bf C}+3\ln(2))}{\pi^2}+\frac{4\left(-\frac{\zeta(3)}{8}+\ln(2)-\frac{\pi^2}{24}\right)}{\pi^2}\, . 
\end{equation}
We have also introduced the Rieman zeta function $\zeta(x)$. Following the same steps as discussed before, we obtain for the large momentum part
\begin{equation}
\label{app1.0.7} \delta_2 (t) = - \frac{1}{8\pi^2 t} \int_1^\infty dp \left\{ \sum_{k=1}^\infty \frac{2}{2k (2k-1)} p^{-2k-2} + 8t \sum_{k=1}^\infty \frac{2}{2k (2k-1)} p^{-2k} e^{8tp^2} Ei(-8tp^2) \right\} \, .
\end{equation}
The transformation $p^2 \rightarrow p$, the expansion of the exponential and an integration by parts yield 
\begin{eqnarray}
&& \delta_2 (t) = - \frac{1}{8\pi^2 t} \sum_{k=1}^\infty \frac{2}{2k (2k-1)(2k+1)} + \frac{1}{\pi^2 } \sum_{k=1}^\infty \frac{2}{2k (2k-1)} \sum_{n=0}^\infty \frac{(8t)^n}{n!} {\Bigg[} \frac{Ei(-8t)}{\left(2n-2k+1\right)} \nonumber\\
\label{app1.0.8} && \hspace*{10.8cm} + 2 \int_1^\infty dp \frac{p^{2n-2k}}{\left(2n-2k+1\right)} e^{-8tp^2}{\Bigg]} \, .
\end{eqnarray}
By expanding the exponential integral the relevant contributions for the asymptotics are identified to come from the $n=0$ term. Then we find
\begin{equation}
\label{app1.0.9} \delta_2 (t\rightarrow 0) =  \frac{1-2\ln(2)}{8\pi^2 t} - \frac{1}{\pi^2} \left\{ \sum_{k=1}^\infty \frac{2\left({\bf C}+\ln(8t)\right)}{2k (2k-1)^2 } - \sum_{k=1}^\infty \frac{4}{2k (2k-1)^3 } \right\} \, .
\end{equation}
Performing the sums exactly the result reads as 
\begin{equation}
\label{app1.0.13} \delta_2 (t\rightarrow 0) =  \frac{1-2\ln(2)}{8\pi^2 t} - \frac{2 \left(\frac{\pi^2}{8}-\ln(2)\right)}{\pi^2}\ln(t)- \frac{2 \left(\frac{\pi^2}{8}-\ln(2)\right)\left({\bf C}+3\ln(2)\right)}{\pi^2}-\frac{8\ln(2)-\pi^2+7 \zeta (3)}{2\pi^2}\, .
\end{equation}
One can combine the results for $\delta_1(t\rightarrow 0)$ [Eq.(\ref{app1.0.12})] and $\delta_2(t\rightarrow 0)$ [Eq.(\ref{app1.0.13})] to find the asymptotic result for $\delta(t\rightarrow 0)$ [Eq.(\ref{0.34})].
\section{Calculation of the second order exchange term}

After the substitution ${\bf k}^\prime = ({\bf p}-{\bf k} -{\bf q})$ Eq.(\ref{0.7}) may be rewritten as
\begin{eqnarray}
\label{app2.1} \epsilon_{ex} (t) & = & - \frac{2t}{\pi^5} \int_{|q_z+p_z+k_z|<1 \atop{|q_z+p_z|>1}} d{\bf q} \int_{|p_z|<1} d{\bf p} \int_{|p_z+k_z|>1} d{\bf k} \frac{1}{{\bf q}^2} \frac{1}{{\bf k}^2} \frac{1}{q_z k_z}  \nonumber\\
&&  \times \exp\left[-2t\left( {\bf p}_\rho^2 + ({\bf q}_\rho+{\bf p}_\rho+{\bf k}_\rho)^2+({\bf q}_\rho+{\bf p}_\rho)^2+({\bf p}_\rho+{\bf k}_\rho)^2\right) \right] \, .
\end{eqnarray}
The Gaussian integrals can be done immediately. The analysis of the conditions on the region of integration gives then
\begin{eqnarray}
\label{app2.2} \epsilon_{ex} (t) & = & - \frac{2}{\pi^2} \int dq_\rho q_\rho \int dk_\rho k_\rho \int_{-1}^1 dp_z \left( \int_{1-p_ z}^2 dk_z \int_{-1-p_z-k_z}^{-1-p_z} dq_z + \int_{2}^\infty dk_z \int_{-1-p_z-k_z}^{1-p_z-k_z} dq_z\right) \nonumber\\
& \times &  \frac{1}{{\bf q}^2} \frac{1}{{\bf k}^2} \frac{1}{q_z k_z} \exp\left[-2t\left(q_\rho^2 + k_\rho^2\right) \right] \, .
\end{eqnarray}
Performing the $p_z$ and $q_z$ integration we find 
\begin{eqnarray}
\label{app2.3} \epsilon_{ex} (t) & = & - \frac{1}{8\pi^2} \int dq_\rho \int dk_\rho \frac{e^{-2q_\rho}}{q_\rho} \frac{e^{-2k_\rho}}{k_\rho} {\Bigg[}  2 \int_2^\infty dk_z \ln\left(1+\frac{q_\rho}{k_z^2 t} \right) \left\{ \ln\left(1+\frac{k_\rho}{(2-k_z)^2t}\right) - \ln\left(1+\frac{k_\rho}{k_z^2t}\right)  \right\} \nonumber\\
& - &  \int_0^2 dk_z  \ln\left(1+\frac{q_\rho}{k_z^2 t} \right) \ln\left(1+\frac{k_\rho}{(2-k_z)^2t}\right) {\Bigg]} \, .
\end{eqnarray}
We split up the integral into two contributions $\epsilon_{ex} (t) = I_1(t) + I_2(t)$ with
\begin{eqnarray}
I_1(t) = - \frac{1}{8\pi^2} \int dq_\rho \int dk_\rho \frac{e^{-2q_\rho}}{q_\rho} \frac{e^{-2k_\rho}}{k_\rho} {\Bigg[} & 2 & \int_0^\infty  dk_z  \ln\left(1+\frac{q_\rho}{k_z^2 t} \right) \left\{ \ln\left(1+\frac{k_\rho}{(2-k_z)^2t}\right) - \ln\left(1+\frac{k_\rho}{k_z^2t}\right)  \right\}  \nonumber\\
\label{app2.112} & - & \int_0^2 \ln\left(1+\frac{q_\rho}{k_z^2 t} \right) \ln\left(1+\frac{k_\rho}{(2-k_z)^2t}\right) {\Bigg]}
\end{eqnarray}
and
\begin{equation}
\label{app2.12} I_2(t) = - \frac{1}{8\pi^2} \int dq_\rho \int dk_\rho \frac{e^{-2q_\rho}}{q_\rho} \frac{e^{-2k_\rho}}{k_\rho}   \int_0^2  dk_z  \ln\left(1+\frac{q_\rho}{k_z^2 t} \right) \left\{ 2 \ln\left(1+\frac{k_\rho}{k_z^2t}\right) - 2 \ln\left(1+\frac{k_\rho}{(2-k_z)^2t}\right)  \right\}  \, .
\end{equation}
First we concentrate on $ I_1(t)$, where after the substitutions $k_\rho/k_z^2 \rightarrow k_\rho$, etc., the $k_z$ integration is again of the Gaussian form. After the transformations $ k = q_\rho k_\rho/(q_\rho + k_\rho) $ and $ q = k_\rho/(q_\rho + k_\rho) $ we have
\begin{equation}
\label{app2.7} I_1(t) = - \frac{1}{8\pi^{\frac{3}{2}}} \int_0^1 dq \int_0^\infty dk  \frac{e^{-4k}-1}{k^{\frac{3}{2}} \sqrt{q(1-q)}} \ln\left(1+\frac{k}{2qt}\right) \ln\left(1+\frac{k}{2(1-q)t}\right) \, .
\end{equation}
Now we investigate this integral in the limit of small $t$ and find for the asymptotic behavior a logarithmic divergence. Noting that
\begin{eqnarray}
\label{app2.8} - \frac{1}{8\pi^{\frac{3}{2}}} \ln^2 (t) \int_0^1 dq \int_0^\infty dk  \frac{e^{-4k}-1}{k^{\frac{3}{2}} \sqrt{q(1-q)}} & = &\frac{1}{2} \ln^2 (t)   \, , \nonumber\\ 
\label{app2.9}  \frac{1}{4\pi^{\frac{3}{2}}} \ln(t)  \int_0^1 dq \int_0^\infty dk  \frac{e^{-4k}-1}{k^{\frac{3}{2}} \sqrt{q(1-q)}} \ln\left( \frac{k}{2q}\right) & = & (3\ln(2)-2+{\bf C}) \ln(t) \, ,\nonumber\\
 - \frac{1}{8\pi^{\frac{3}{2}}} \int_0^1 dq \int_0^\infty dk  \frac{e^{-4k}-1}{k^{\frac{3}{2}} \sqrt{q(1-q)}} \ln\left( \frac{k}{2q}\right) \ln\left( \frac{k}{2(1-q)}\right) & = & {\bigg(}4 - 2{\bf C} + \frac{{\bf C}^2}{2} + \frac{\pi^2}{4} + 3 {\bf C } \ln(2) \nonumber\\
\label{app2.10} & - & 6 \ln(2) + \frac{9}{2}\ln^2(2)-\frac{\pi^2}{12} {\bigg)} \,  
\end{eqnarray}
and neglecting all terms that vanish as $t\rightarrow 0$, this contribution becomes
\begin{equation}
\label{app2.11} I_1(t\rightarrow 0)  =   \left(\frac{1}{2} \ln^2(t) + (3\ln(2)-2+{\bf C}) \ln(t) + 4 - 2{\bf C} + \frac{{\bf C}^2}{2} + \frac{\pi^2}{6} + 3 {\bf C } \ln(2) - 6\ln(2) + \frac{9}{2}\ln^2(2) \right) \, .
\end{equation}
We now find an appropriate approximation scheme for the calculation of $I_2$. We may perform the identical $k$ and $q$ integration [see Ref. \cite{Prudnikov}, p. 530, Eq.(6)] to obtain the result
\begin{eqnarray}
\label{app2.13} I_2(t) & = & - \frac{1}{8\pi^2} \frac{1}{4} \int_0^2 dk_z \left\{\ln^2(2 k_z^2 t) + 2 {\bf C} \ln(2 k_z^2 t) +{\bf C}^2 + \frac{\pi^2}{2}- 2 \sum_{n=1}^\infty \frac{2n^{-1}+\psi(n)- \ln(2 k_z^2 t)}{n! n} \left( 2 k_z^2 t \right)^n \right\} \nonumber\\
& \times & \left\{2 \left(k_z^2 \leftrightarrow k_z^2\right) - 2 \left(k_z^2 \leftrightarrow (2-k_z)^2 \right) \right\} \, .
\end{eqnarray}
We find this may be exactly rewritten as
\begin{equation}
\label{app2.22} I_2(t)  =  C1 \ln^2(t) + C2 \ln(t) + C3 +  \sum_{n=1}^\infty \left(C4(n) \ln^2(t)+C5(n) \ln(t)+C6(n)\right) t^n 
\end{equation}
with the constant coefficients
\begin{eqnarray}
\label{app2.31} C1 & = & \frac{1}{16 \pi^2} \int_0^2 dk_z 4 \ln\left(2 k_z^2\right) \left( \ln\left(2[2-k_z]^2\right)-\ln \left(2 k_z^2\right)  \right) \nonumber\\
& = & - \frac{1}{3} \, ,\nonumber\\
\label{app2.32} C2 & = & \frac{1}{16 \pi^2} \int_0^2 dk_z {\Bigg(} 2 \ln(2 k_z^2) \left(2 {\bf C} + \ln(2 k_z^2) \right) \left( \ln(2 [2-k_z]^2)- \ln(2 k_z^2)\right) + 2 \ln(2 k_z^2) \\
& \times & \left(\ln^2(2 [2-k_z]^2)+2 {\bf C} \ln(2 [2-k_z]^2)-\ln^2(2 k_z^2)- 2 {\bf C} \ln(2 k_z^2)\right) {\Bigg)}\nonumber\\
& = & \frac{4}{3}-\frac{2}{3} {\bf C}+\frac{8}{\pi^2} \zeta(3)-2 \ln(2) \, ,\nonumber\\
\label{app2.33} C3 & = & \frac{1}{16 \pi^2} \int_0^2 dk_z \ln(2 k_z^2) \left(2 {\bf C} + \ln(2 k_z^2) \right) \left(\ln^2(2 [2-k_z]^2)+2 {\bf C} \ln(2 [2-k_z]^2)-\ln^2(2 k_z^2)- 2 {\bf C} \ln(2 k_z^2)\right) \nonumber\\
& = & -3 \ln^2(2)+\left(4-2 {\bf C}\right) \ln(2)-\frac{{\bf C}^2}{3}-\frac{8}{3}-\frac{2}{90}\pi^2 +\frac{4}{3} {\bf C} + \frac{\zeta(3)}{\pi^2} \left(24 \ln(2)+8 {\bf C}-16\right) \, .\nonumber\\
\end{eqnarray}
We neglect higher-order terms in this calculation. Thus we get the asymptotic expansion
\begin{eqnarray}
 I_2(t \rightarrow 0) & = & - \frac{1}{3} \ln^2(t)+\left(\frac{4}{3}-\frac{2}{3} {\bf C}+\frac{8}{\pi^2} \zeta(3)-2 \ln(2)\right) \ln(t)-3 \ln^2(2)+\left(4-2 {\bf C}\right) \ln(2)-\frac{{\bf C}^2}{3} \nonumber\\
\label{app2.30} & & -  \frac{8}{3}-\frac{2}{90}\pi^2 +\frac{4}{3} {\bf C} + \frac{\zeta(3)}{\pi^2} \left(24 \ln(2)+8 {\bf C}-16\right)\, . 
\end{eqnarray}
Summing $I_1$ and $I_2$, we have the final expression for the second-order exchange energy at small values of the filling parameter $t$ given in Eq.(\ref{0.22}).
\end{appendix}

\end{document}